\documentclass[prl,twocolumn,showpacs,superscriptaddress]{revtex4}
\usepackage{amsmath,amssymb,graphicx,cases}

\begin{document}
\title{Screening of a hypercritical charge in graphene}

\author{M. M. Fogler}


\affiliation{Department of Physics, University of California San
Diego, La Jolla, 9500 Gilman Drive, California 92093}

\author{D. S. Novikov}
\author{B. I. Shklovskii}

\affiliation{W. I. Fine Theoretical Physics Institute, University of
Minnesota, Minneapolis, Minnesota 55455}

\date{\today}

\begin{abstract}

Screening of a large external charge in graphene is studied. The charge is
assumed to be displaced away or smeared over a finite region of the
graphene plane. The initial decay of the screened potential with
distance is shown to follow the 3/2 power. 
It gradually changes to the Coulomb law outside of a hypercritical core
whose radius is proportional to the external charge.

\end{abstract}



\maketitle

Recent discovery of graphene -- a two-dimensional (2D) form of
carbon~\cite{Novoselov_04} -- brought an exciting link between
solid-state physics and quantum electrodynamics (QED). The half-filled
$\pi$-band of graphene has a relativistic massless Dirac spectrum
$\epsilon = \pm \hbar v |{\bf k}|$ where $\epsilon>0$ for the electrons
and $\epsilon<0$ for holes, ${\bf k}$ is the deviation of the
quasi-momentum from the Brillouin zone corner, and $v \approx 10^6\,
{\rm m}/{\rm s}$. The role of the fine-structure constant is played by
the dimensionless parameter
\begin{equation}
\alpha = {e^2}/{\kappa \hbar v} \,, \quad e^2/\hbar v \approx 2.2 \,,
\label{eq:alpha}
\end{equation}
where $\kappa$ is the dielectric constant at the interface of
substrate and vacuum. For conventional SiO$_2$ substrates
$\kappa\approx 2.4$; hence, Coulomb interaction is strong, $\alpha
\sim 1$.

In this work we consider the problem of screening of a Coulomb potential
$V_0 = e Z /\kappa r$ that can be induced in graphene by a group of
charged impurities in the substrate, by a nearby gate, or by a cluster
of dopants. This problem is important for a number of properties of
graphene nanostructures, including transport~\cite{Ando_06, Nomura_07,
Hwang_07, Adam_07}, local gating~\cite{Ozyilmaz_xxx, Huard_07,
Williams_07, Martin_xxx}, controlled doping~\cite{Ohta_06, Uchoa_xxx},
and chemical sensing~\cite{sensors}. Not surprisingly, it has attracted
much attention~\cite{DiVincenzo_84, Gonzales_94, Ando_06, Katsnelson_06,
Hwang_07, Adam_07, Shytov_xxx, Novikov_xxx, Pereira_xxx, Biswas_xxx}. In
particular, it has been noted \cite{Novikov_xxx, Shytov_xxx,
Pereira_xxx} that at half-filling the problem of a Coulomb charge in
graphene has an interesting parallel with that of a hypothetical
supercritical atom with $Z > \hbar c / e^2 \approx 137$. For such an
atom, the standard solution~\cite{DarwinGordon} of the Dirac equation
breaks down and a physically acceptable atomic structure is obtained
only after accounting for a finite radius of the
nucleus~\cite{Pomeranchuk_45}. This structure is
characterized by a vacuum reconstruction: a certain number of electrons
is spontaneously created (liberating positrons), they bind to the
nucleus, and render it subcritical. In graphene the critical charge
\cite{Khalilov_98, Shytov_xxx, Novikov_xxx, Pereira_xxx} $Z_c \simeq 1 /
2\alpha$ is much smaller than in QED; hence, solid-state analogs of
supercritical atoms may be realizable even at $Z \sim 1$.

According to all prior investigations, screening properties of an
undoped graphene resemble those of a dielectric: the screened potential
$V$ of a supercitical charge has been argued not to deviate much from
the Coulomb law,
\begin{equation}
V(r) = \frac{e}{\kappa r} \frac{F(r)}{2 \alpha^2}\,.
\label{eq:F}
\end{equation}
Here $F(r)$ is a slow logarithmic function. Such a conclusion follows
from the standard linear response theory --- Random Phase Approximation
(RPA)~\cite{Gonzales_94, Ando_06} --- and was supposedly confirmed by
calculations within the Thomas-Fermi (TF) method~\cite{DiVincenzo_84,
Katsnelson_06, Shytov_xxx} that is able to go beyond the linear
response. Below we re-examine these conclusions for the case of a
hypercritical charge $Z\gg 1$, which lets itself to a controlled
treatment and adds new physics. Without loss of generality we assume
that the external charge attracts electrons, $Z > 0$.

Since it was not always made clear previously, we emphasize that the
problem is ill-defined unless one explicitly regularizes the strong
Coulomb singularity at the origin. This is as crucial as introducing a
finite size of a nucleus in QED. Therefore, the charge $Z$ must be
either displaced away from graphene plane by some distance $d$ or spread
over the area of some radius $r_0$ in this plane. In order to deal
exclusively with Dirac fermions the smearing parameter $\max\{d, r_0\}$
must exceed $a \alpha \sqrt{Z}$ where $a = 2.5\,\text{\AA}$ is the graphene
lattice constant; otherwise, the quasiparticle energy shift due to the
potential $V$ would exceed the modest energy separation $4\,{\rm eV}
\sim e^2 / a$ of the Dirac point and the nearest
$\sigma$-bands~\cite{Saito_92}. These other bands would then also need
to be included, leading one to a three-dimensional (3D) problem that has
little to do with special properties of graphene.

Our main result is that the induced 2D electron density
and the screened potential have the form
\begin{equation}
n(r) \simeq \frac{1}{4\pi \alpha^2} \frac{r_1}{r^3}\,, \:\:
V(r) \simeq \frac{e}{2 \alpha^2 \kappa}
     \sqrt{\frac{r_1}{r^3}}\,,
\:\: r_1 \equiv 2 \alpha^2 Z d
\label{eq:3_2-law}
\end{equation}
in the range of distances $\max\{d, r_0\} \ll r \ll r_1$. 
Based on electrostatics, the law (\ref{eq:3_2-law}) is robust and universal.
How does then one reconcile it with Eq.~(\ref{eq:F})?
As we clarify below, the situation is
as follows. In the strongly interacting case, $\alpha\sim 1$,
Eq.~(\ref{eq:3_2-law}) controls the entire supercritical core, i.e., the
circle around the origin where the net charge exceeds $Z_c$. This fact
has eluded previous studies. However, if $\alpha$ is small, the domain of
validity of Eq.~(\ref{eq:3_2-law}) narrows down, opening up a window
where Eq.~(\ref{eq:F}) is realized. Although current
experiments are not in this regime, small $\alpha$ can be achieved using
large $\kappa$ substrates, e.g., HfO$_2$~\cite{Ozyilmaz_xxx} or simply
liquid water, $\kappa \sim 80$.


The three-line derivation of Eq.~(\ref{eq:3_2-law}) can be given if, as
discussed above, the charge $Z$ is point-like but removed from the
graphene plane~\cite{DiVincenzo_84} by an appropriate distance $d$. The
key idea is that if we treat the graphene sheet as a perfect metal, then
classical electrostatics dictates that the induced charge
density is given by $n = n_\text{cl}$, where
\begin{equation}
n_\text{cl}(r) = \frac{1}{2 \pi} \frac{Z d}{(r^2 + d^2)^{3/2}}
               = \frac{1}{4\pi \alpha^2}
                 \frac{r_1}{(r^2 + d^2)^{3/2}}\,.
\label{eq:n_charge}
\end{equation}
At $r \gg d$ we get the first formula in Eq.~(\ref{eq:3_2-law}).
To derive $V(r)$ we employ the TF approximation,
\begin{equation}
\mu[n(r)] - e V(r) = 0.
\label{eq:TFA}
\end{equation}
Combined with the formula for the chemical potential,
\begin{equation}
\mu(n) = {\rm sign}(n)\sqrt{\pi} \hbar v |n|^{1/2}\,,
\label{eq:mu}
\end{equation}
specific for the 2D Dirac spectrum, it yields the second formula in
Eq.~(\ref{eq:3_2-law}), concluding the derivation. The rest of our paper
is needed mainly to explain why the above reasoning is correct, why
Eq.~(\ref{eq:3_2-law}) is completely general rather than restricted to
the case of a remote charge, and finally, where the room may still exist
for the differing predictions for $n$ and $V$ advocated in
Refs.~\cite{DiVincenzo_84, Katsnelson_06, Shytov_xxx}.

First, let us clarify why it was legitimate to approximate the density
response of graphene --- a complicated quantum system --- simply by that
of an ideal metal. The reason is this. At $r \ll r_1$ the local screening
length $r_s = ({\kappa}/\,{2 \pi e^2}) ({d \mu}/{d n}) \sim
\alpha^{-1} |n|^{-1 / 2}$ is much smaller than the characteristic scale
$\max\{r,d\}$ over which the potential $V(r)$, or equivalently, the
effective background 2D charge density $n_\text{cl}(r)$ vary. Therefore,
the unscreened charge density, $\sigma(r) \equiv n_\text{cl}(r) - n(r)$,
is smaller than the background one, $n_\text{cl}(r)$, by some large
factor related to the ratio of $r_s$ and $\max\{r,d\}$. [The precise
relation is expressed by Eqs.~(\ref{eq:n_0}) and (\ref{eq:n_far})
below.]

The next step is to explain why or rather \emph{where\/} the TF
approximation can be trusted. This is determined by the conditions that
$\max\{r, d\}$ exceeds the local Fermi wavelength $\lambda_F(r) \sim
n^{-1/2}(r)$. For $\alpha\sim 1$ we can use $n(r)$ from
Eq.~(\ref{eq:3_2-law}) to write this condition as $r \lesssim r_2 = Z
d$. Thus, for $\alpha\sim 1$ the domains of validity of the TF and the
perfect screening approximations coincide, $r_1 \sim r_2$. At $r \ll
r_2$ all corrections to Eq.~(\ref{eq:3_2-law}), both smoothly varying
with $r$ and Friedel oscillations~\cite{Cheianov_06, Shytov_xxx} are
subleading.

Let us briefly discuss the nature of screening at $r > r_2$ where the TF
approximation breaks down. Define $Q(r)$ to be the net effective charge
inside the circle of radius $r$,
\begin{equation}
\label{eq:Q_def}
Q(r) \equiv \int\limits_0^r\!
            2 \pi \sigma(r^\prime) r^\prime d r^\prime\,.
\end{equation}
At $r = r_2$, $Q$ drops to a number of the order of the critical one
$Z_c \sim {1}/{2\alpha}$. Consideration of screening now requires a
detailed analysis of the eigenstates of the Dirac
equation~\cite{Shytov_xxx, Novikov_xxx, Pereira_xxx} in the potential
created by the charge $Q(r_2)$. According to Ref.~\cite{Shytov_xxx},
some amount of charge, in fact, exactly the critical one remains
unscreened: $Q(\infty) = Z_c$. The saturation of $Q$ at this value
occurs near a certain $r = r_\ast$. However for $\alpha \sim 1$,
$r_\ast$ and $r_2$ must coincide up to a factor of the order of unity.
Thus, Eq.~(\ref{eq:3_2-law}) governs the entire supercritical core
except perhaps a non-parametrically wide outer region $r \sim r_2$ where
a more complicated dependence~\cite{Shytov_xxx} may apply.
At even larger distances the potential $V(r)$ follows the RPA prediction
\begin{equation}
V(r) \simeq e Z_c \,/\, {\varepsilon r}\,,
\quad r \gg r_2\,,
\label{eq:V_RPA}
\end{equation}
where $\varepsilon = \kappa [1 + ({\pi}/{2}) \alpha]$ is the RPA
dielectric constant~\cite{Gonzales_94, Ando_06}. A more careful
examination of the behavior of $V(r)$ at such $r$ requires accounting
for the infrared renormalization of $\alpha$ (which enters
$\varepsilon$) \cite{Gonzales_94, Son_07, Biswas_xxx}, that is not
directly related to the problem at hand.


Let us return to the analysis of the supercritical region and show how
to refine our results by computing corrections to
Eq.~(\ref{eq:3_2-law}). For this we complete the set of the TF
Eqs.~(\ref{eq:TFA}) and (\ref{eq:mu}) by adding another one for $V(r)$:
\begin{equation}
\frac{\kappa}{e}V = \!\int\!
\frac{d^2 {\bf r}^\prime \sigma({\bf r}^\prime)}{|{\bf r} - {\bf r}^\prime|}
= \int\limits_0^\infty\! d q J_0(q r) \tilde\sigma(q)
= \int\limits_0^r\! \frac{g(s) d s}{\sqrt{r^2 - s^2}}
\label{eq:V_from_n}
\end{equation}
where $J_0(z)$ is the Bessel function~\cite{Gradshteyn_Ryzhik} and
$\tilde\sigma$, aptly parametrized by $\tilde\sigma(q) =
\textstyle{\int_0^\infty} d s\, g(s) \cos q s$ ~\cite{Sneddon}, is the
2D Fourier transform of $\sigma$. 
Inverting the last equation of (\ref{eq:V_from_n}), 
we get
$g(u) = (2 \kappa /\pi e)(d / du) \int_0^u V(s) s ds/\sqrt{u^2-s^2}$
and
\begin{equation}
Q(r) = Q(\infty) - \frac{2}{\pi} \frac{\kappa}{e^2}
\int\limits_r^\infty\! \frac{u d u}{\sqrt{u^2-r^2}}
\frac{d}{d u}
\int\limits_0^u \! \frac{e V(s) s d s}{\sqrt{u^2-s^2}}
\,.
\label{eq:Q}
\end{equation}
The leading correction to the perfect screening can be obtained by
substituting $\mu\left[ n_\text{cl}(s)\right]$ in lieu of $eV(s)$,
cf.~Eq.~(\ref{eq:TFA}). The resultant expression is cumbersome, and so
we quote only the limiting forms:
\begin{numcases}{
\frac{\sigma(r)}{n_\text{cl}(r)} \simeq}
        -{\Gamma^2 (5/4)}
                 \sqrt{\frac{8 d}{\pi r_1}}\,,&
                 $r \ll d$,
\label{eq:n_0}\\
        \frac{16}{\pi^2}\, {\Gamma^4 (5/4)}
         \sqrt{\frac{r}{r_1}}\,,&
                 $d \ll r \ll r_1$, \quad
\label{eq:n_far}
\end{numcases}
where the Gamma-function~\cite{Gradshteyn_Ryzhik} $\Gamma(5/4)\approx
0.906$. In agreement with the above physical argument, the deviation
from the perfect screening at all $r \ll r_1$ is small.

\begin{figure}
\centerline{
\includegraphics[width=3.0in]{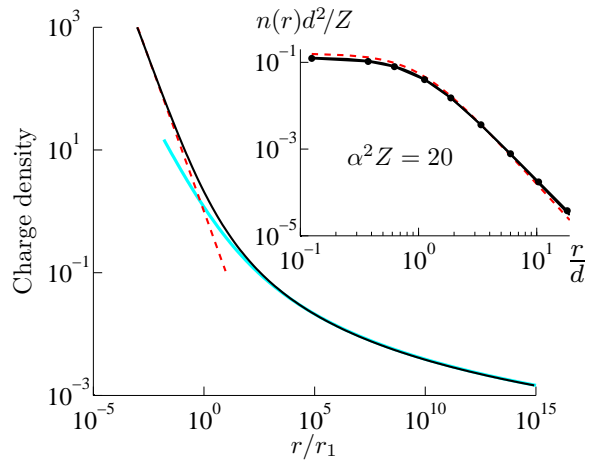}
}

\caption{ \label{Fig:results} (Color online) Main panel: Density profile
in the limit of $d \to 0$ at fixed $r_1$. The quantity plotted on the
vertical axis is $4 \pi \alpha^2 r_1^2 n(r) = F^2(r) r_1^2 / r^2$. The
thin black line is the numerical solution of
Eq.~(\ref{eq:TF_Katsnelson}); the red dashed line is the perfect
screening, $F = r_1 / r$; the thick cyan line is for $F$ from
Eq.~(\ref{eq:F_eq}) with $c = 0.6$. Inset: expanded view of $n(r)$
inside the hypercritical core. The thin black line and the red dashed
line have the same meaning as before; the dots correspond to an
analytical formula whose limits are given by Eqs.~(\ref{eq:n_0}) and
(\ref{eq:n_far}).}

\end{figure}

These analytical predictions were verified by numerical simulations. To
this end we solved the TF equations~(\ref{eq:TFA})--(\ref{eq:V_from_n})
inside a finite square of the 2D plane. The integrals were replaced by
discrete sums over a uniform $256 \times 256$ grid defined therein and
the periodic boundary conditions were imposed. The solution for $n(r)$
and $V(r)$ was found by a standard iterative method, using
underrelaxation to ensure convergence. As shown in the inset of
Fig.~\ref{Fig:results}, the analytical and the numerical results agree
extremely well for a suitably large hypercritical charge $\alpha^2 Z =
20$.


{\it Small--$\alpha$ regimes.---} Let us now show that
Eq.~(\ref{eq:F}) can be reconciled with our theory under the
condition $\alpha \ll 1$, i.e., $\kappa \gg 1$. In this case
there is a gap between the above defined characteristic lengthscales
$r_1$ and $r_2$. This gap is filled by an additional
regime where the TF approximation regime is still valid but the
screening is ineffective.

To see that consider first moderately small $\alpha$, such that
$1/\sqrt{Z} \ll \alpha \ll 1$. Since the screening is weak,
Eq.~(\ref{eq:Q}) is no longer convenient. Instead, the derivation of $V$
and $n$ can be done along the lines of Ref.~\cite{Katsnelson_06} but
with several important refinements. First, we trade the two last equations
of~(\ref{eq:V_from_n}) for
\begin{equation}
\frac{\kappa}{e} V(r)
= \int\limits_0^\infty \!
 \frac{4 r^\prime d r^\prime}{r^\prime + r}
 K\!\left(\frac{2 \sqrt{r r^\prime}}{r + r^\prime}\right)
 [n_\text{cl}(r^\prime) - n(r^\prime)]
\,,
\label{eq:V_from_n_II}
\end{equation}
where $K(z)$ is the complete elliptic integral of the first
kind~\cite{Gradshteyn_Ryzhik}. Next, we treat
Eq.~(\ref{eq:F}) as the definition of yet unknown function $F$
and use Eqs.~(\ref{eq:TFA}) and (\ref{eq:mu}) to obtain
\begin{equation}
n(r) = {F^2(r)} \,/\, {4 \pi \alpha^2 r^2}\,.
\label{eq:n_Katsnelson}
\end{equation}
Taking the limit $d \to 0$ at fixed $r_1$, we get the equation
\begin{equation}
F(t) = \int\limits_{-\infty}^\infty \! d u
[\theta(t - u) + \phi(u - t)] [e^{-u} - F^2(u)]\,,
\label{eq:TF_Katsnelson}
\end{equation}
where $t = \ln ({r}/{r_1})$ and $\theta(t)$ is the unit step-function.
Function $\phi(t)$, defined by Eq.~(12) of Ref.~\cite{Katsnelson_06},
has the following properties: it is a logarithmically divergent at $t =
0$, is exponentially small at $|t| \gg 1$, and satisfies $\int \phi(t) d
t = \ln 4$. It is easy to see that at large negative $t$, we must have
perfect screening, $F^2(t) \simeq e^{-t}$. The asymptotic behavior
of $F$ at large $t > 0$ can be deduced by replacing $\phi(t)$ with
$(\ln 4) \delta(t)$~\cite{Comment_on_ln4}. After this, we can
differentiate the integral equation~(\ref{eq:TF_Katsnelson}) to get
\begin{equation}
F^{-1} - (2 \ln 4) \ln F = \ln(r / r_1) + c\,,
\quad c = \text{const}\,,
\label{eq:F_eq}
\end{equation}
where we returned to the original linear coordinate $r$. The direct
numerical solution of Eq.~(\ref{eq:TF_Katsnelson}) shows in excellent
agreement with Eqs.~(\ref{eq:F}) and Eq.~(\ref{eq:F_eq}) if the
constant $c$ is set to $0.6$, see Fig.~\ref{Fig:results}. At
$r \sim r_1$, this solution crosses over to the
strong screening regime, Eq.~(\ref{eq:n_far}).

The range of $r \ll r_2$ where Eq.~(\ref{eq:F_eq}) is
valid is again determined by the condition $r \gg n^{-1/2}(r)$, which
yields
\begin{equation}
                   r_2 \sim r_1 \exp(1 /\, 2 \sqrt{\pi}\, \alpha).
\label{eq:r_2}
\end{equation}
At $r \gg r_2$ the screened potential is given by Eq.~(\ref{eq:V_RPA}).

Consider even smaller $\alpha$, such that $1/Z \ll \alpha \ll
1/\sqrt{Z}$. Here the Coulomb interactions are so weak that the smearing
of the external charge is no longer necessary: the ``dangerous'' region
$r < a \alpha \sqrt{Z}$ is smaller than the lattice constant. In
addition, the domain of the perfect screening, $r < r_1$, which is the
region of validity of Eq.~(\ref{eq:3_2-law}) disappears. (Weak
interactions entail poor screening.) In this case $c \to (2 \alpha^2
Z)^{-1} + \ln (r_1 / a)$, so that the solution
\begin{equation}
V(r) \simeq \frac{e Z}{\kappa r}
\frac{1}{1 + 2 \alpha^2 Z \ln (r / a)}\,,
\label{eq:V_Katsnelson}
\end{equation}
advocated in Ref.~\cite{Katsnelson_06} actually applies, at $\ln(r / a)
< 1 / \alpha$.

{\it In-plane charge.---} In the concluding part of the paper we wish to
return to the structure of the hypercritical core and to show that
Eq.~(\ref{eq:3_2-law}) remains valid if the charge $Z$ resides within
the 2D plane. To gain some intuition consider first an artificial
scenario where the external charge is highly localized yet the
$\sigma$-bands of graphene can be neglected. In this case the maximum
possible electron density (relative to that of the half-filled $\pi$-band)
is $n_{\rm max} = 2 / \sqrt{3} a^2$. This density is indeed reached
at $r$ smaller than some radius $b$ as a result of attraction of electrons
to the hypercritical charge $Z$. At $r > b$, electron density is
gradually decreases, which can be thought of appearance of ``holes'' at
the top of the conduction band.

Incidentally, the charge profile of these holes within the perfect
screening approximation is known exactly. It can be read off the results
of Ref.~\cite{Deruelle_92} where the structure of a depletion region
around a disk of a negative charge in a semiconductor was studied. For a
high density of the external charge these authors found that $b = (Z / 2
\pi n_\infty)^{1/2}$, where $n_\infty$ is the uniform electron density
far away form the depletion. They also found~\cite{Deruelle_92} that the
density profile at large $r$ is given by $n(r) = n_\infty - (Z b / {2
\pi r^3})$ at $r \gg b$. Adopting these results to our problem, we get
\begin{equation}
       n(r) = Z b / {2 \pi r^3}\,,\quad
       r \gg b = (Z / 2 \pi n_{\max})^{1/2}\,,
\label{eq:n_Deruelle}
\end{equation}
leading to Eq.~(\ref{eq:3_2-law}) with
$r_1 = 2 \alpha^2 Z b \sim a Z^{3 / 2}$ for $\alpha \sim 1$.

Consider now a more realistic setup where the external charge $n_{\rm
ext}({ r})$ is distributed over a disk of radius $r_0 \gg a \sqrt{Z}$.
Then $n \leq n(0) \sim Z / \pi r_0^2 \ll n_{\max}$, so that
$\sigma$-bands bands can indeed be disregarded. Let us
show that
\begin{equation}
 n(r) \sim \frac{Z r_s}{2 \pi r^3}\,,\quad
 r_0 \ll r \ll r_1 = 2 \alpha^2 Z r_s\,,
\label{eq:sigma_antidot}
\end{equation}
where the screening length $r_s \sim 1 / \alpha \sqrt{n(0)}$.

Based on the near-perfect screening framework used in the first part of
the paper [and justified {\it a posteriori\/} by
Eq.~(\ref{eq:sigma_antidot})] we can claim that $V(r)$ is substantial
only in the region $r < r_0$ and is greatly reduced at $r > r_0$. This
implies that the Fourier transform of $V$ is nearly
wavevector-independent over a range of $q$,
%
\[
\widetilde{V}(q) = c_1 Z e r_s / \kappa + {\cal O}(Z^{1/2})\,,
                 \quad r_1^{-1} \ll q \ll r_0^{-1}\,.
\]
%
The first term, with $c_1 \sim 1$, follows from Eq.~(\ref{eq:TFA}). In
turn, the Fourier transform of the charge density, $
\tilde\sigma(q) = \tilde{n}_{\rm
ext}(q) - \tilde{n}(q) = \widetilde{V}(q) / (2 \pi e /
\kappa q)$, that produces this potential behaves as
$\tilde\sigma(q) = c_2 Z r_s q + {\cal O}(Z^{1/2})$, where $c_2 \sim 1$.
After the inverse Fourier transform, the net charge density $\sigma(r)$
is seen to be dominated by the term $-c_2 Z r_s / 2 \pi r^3$ at $r_0 \ll
r \ll r_1$. Since $n_{\rm ext}(r) = 0$ for such $r$, this term is
entirely due to $n$, proving our statement.

In summary, we considered the problem of nonlinear screening of a large
charge by the massless electrons in graphene. The consistent formulation
of the problem requires the charge to be either displaced from the
graphene plane or to be spread over a disk of finite radius $r_0$. In
both cases the screening is nonlinear within a region of a
parametrically large radius $r_1$. In the interval between $r_0$ and
$r_1$ the screened potential decays as $1 / r^{3/2}$. Our results are
relevant for current and future experiments that involve local charging
or doping of graphene. Thus, if small $\alpha$ can be
achieved experimentally, it may be possible to verify the predicted
crossover from Eq.~(\ref{eq:3_2-law}) to (\ref{eq:F}) and
finally to (\ref{eq:V_Katsnelson}) by using scanned probe
techniques~\cite{Martin_xxx}.


We are grateful to L.~Glazman, M.~Katsnelson, L.~Levitov, and A.~Shytov
for useful discussions. D.~N. was supported by the NSF grants DMR
02-37296 and DMR 04-39026. M.~F. thanks the Aspen Center for Physics for
hospitality during the completion of this paper.


\end{document}